\documentclass[aps,showpacs,prep,graphics,twocolumn]{revtex4}
\usepackage{amssymb}
\usepackage[dvips]{graphicx}
\usepackage{amsmath}
\usepackage{bm}
\usepackage{epsfig}

\begin{document}

\title{New effective coupled $F(^{(4)}\!R,\varphi)$ modified gravity from  $f(^{(5)}\!R)$ gravity in five dimensions}

\author{  Jos\'e Edgar Madriz Aguilar \thanks{E-mail address: madriz@mdp.edu.ar} }
\affiliation{ Departamento de Matem\'aticas, Centro Universitario de Ciencias Exactas e ingenier\'{i}as (CUCEI),
Universidad de Guadalajara (UdG), Av. Revoluci\'on 1500 S.R. 44430, Guadalajara, Jalisco, M\'exico.  \\
E-mail: madriz@mdp.edu.ar,  edgar.madriz@red.cucei.udg.mx}

\begin{abstract}
We obtain a new kind of $F(^{(4)}\!R,\varphi)$ modified gravity theory as an effective four-dimensional (4D) theory derived from  $f(^{(5)}\!R)$ gravity in five dimensions (5D). This new theory exhibits a different matter coupling than the one in BBHL theory. We show that the field equations of the Wesson's induced matter theory and of some braneworld scenarios can be obtained as maximally symmetric solutions of the $f(^{(5)}\!R)$ theory. We found criteria for the Dolgov-Kawasaki instabilities for both the $f(^{(5)}\!R)$ and the $F(^{(4)}\!R,\varphi)$ theories. We demonstrate that under certain conditions imposed on the 5D geometry it is possible to interpret the $F(^{(4)}\!R,\varphi)$ theory as a modified gravity theory with dynamical coefficients, making this new theory a viable candidate to address the present accelerating cosmic expansion issue.
\end{abstract}

\maketitle

\vskip .5cm
$f(R)$-gravity, induced matter theory, brane worlds, Dolgov-Kawasaki instabilities.

\section{Introduction}

The acceleration in the expansion of the universe observed since 1998 by the Supernova Cosmology Project \cite{Nova2}, has generated a great quantity of research in gravitation. In the quest of a satisfactory explanation of these phenomena, theoretical physicists have basically followed three lines of reasoning \cite{Re1}. First they search for some new properties of standard gravity models capable to bring out an explanation. Second, they atribute the acceleration to a dark energy component of the universe. However, in view of the problems arising from this idea, related mainly with the nature and origin of dark energy, many other researchers resort to a third class of theories: modified gravity theories. We can find in the literature a plenty of proposals such as scalar-tensor theories \cite{Re2,Re3}, $f(R)$ theories \cite{Re4,Re4p}, DGP gravity \cite{Re5}, braneworld scenarios \cite{Re6,Re7,Re8}, induced matter theory \cite{wbook,rept,Wesson} and modified gravity with dynamical coefficients \cite{Re9}, among many others.\\

During the last decade, $f(R)$ theories have receive a great deal of attention because represent a possibility to address the cosmic accelerating expansion and dark matter issues \cite{Re10}. In order to have a viable $f(R)$ theory there is a minimal criterio, for instance, the theory must reproduce the cosmic dynamics in good agreement with observations and the theory must be free of instabilities. One very common in the matter sector is the Dolgov-Kawasaki instability \cite{Re11}. The consideration of physically different instabilities yields remarkably similar stability conditions \cite{Faraoni1,Re11p}. Ghosts instabilities may be also present \cite{Re12}.\\

With the idea to have viable modified theories of gravity, generalizations of $f(R)$ theories have been proposed \cite{Re12p}. One example of non-minimal $f(R)$ gravity theories are those which exhibe couplings of the scalar curvature with matter, like the Bertolami, Bohmer, Harko and Lobo (BBHL) theory \cite{Re13,Re14} . In these kind of models a fifth force on massive particles appears causing changes in the acceleration law derived in the weak field limit of BBHL theory, in a similar manner than in the acceleration law in MOND models \cite{Re15}.\\

In this letter we derive a new effective coupled $F(^{(4)}\! R,\varphi)$ modified gravity theory from a five dimensional (5D) $f(^{(5)}\! R)$ theory of gravity, where the fifth extra coordinate is considered extended (non-compact). The letter is organized as follows. In section I we give a brief introduction. In section II we obtain the 5D field equations of the theory, together with a formulation of a criterion to avoid the Dolgov-Kawasaki instability in 5D. In adition, we show how the field equations of some braneworld models and of the induced matter theory can be obtained as particular maximally symmetric solutions of the 5D theory. In section III we obtain the induced effective 4D field equations of the $F(^{(4)}\! R,\varphi)$ theory and discuss its Dolgov-Kawasaki instability criterion. Some examples of how to induced a $F(^{(4)}\! R,\varphi)$ from its 5D analog $f(^{(5)}\! R)$ are also included in this section. Finally in section IV we give some final comments. In our conventions latin indices like $a,b,c,$ etc. run from $0$ to $4$, latin indices like $i,j,$ etc., run from $1$ to $3$ and greek indices take values from $0$ to $3$.

\section{Dynamical aspects of  $f(^{(5)}R)$-gravity}

Let us start considering a $f(R)$ theory of gravity in five dimensions described by the action
\begin{equation}\label{p1}
^{(5)}\!{\cal S}=\frac{1}{2\kappa_5}\int d^{5}\!y\sqrt{g_5}\left[f(^{(5)}\!R)+{\cal L}_{m}(g_{ab},\psi)\right],
\end{equation}
being $\kappa_5$ the 5D gravitational coupling, $^{(5)}R$ the 5D Ricci scalar, ${\cal L}_m(g_{ab},\psi)$ a lagrangian density for matter fields denoted by $\psi$ and $g_5$ the determinant of the 5D metric tensor $g_{ab}$. The field equations derived from the action (\ref{p1}) in the metric formalism read
\begin{eqnarray}
&& f_{,R}(^{(5)}\!R)\,^{(5)} \! R_{ab}\, - \,\frac{1}{2}\, f(^{(5)}\!R)\, g_{ab}\, \nonumber\\
\label{p2}
&& -\, [\nabla\!_a\nabla\!_b-g_{ab}\,^{(5)}\Box]\,f_{,R}(^{(5)}\!R)=\kappa_5\,^{(5)}T_{ab},
\end{eqnarray}
where $T_{ab}(\psi)$ is the energy-momentum tensor for matter sources, $\nabla_a$ is the 5D covariant derivative, $^{(5)}\!\Box =g^{ab}\nabla_a\nabla_b$ is the 5D D'Alambertian operator and $f_{,R}$ is denoting derivative with respect to $^{(5)}\!R$.\\

By taking the trace of the eq. (\ref{p2}) we obtain
\begin{equation}\label{p3}
4\,^{(5)}\Box f_{,R}+f_{,R}\,^{(5)}\!R-\frac{5}{2}f=\kappa_5\,^{(5)}T
\end{equation}
with $^{(5)}T\equiv g^{AB}\,^{(5)}T_{AB}$ being the trace of $^{(5)}T_{ab}$. This equation will allow us to study some important aspects of the $f(^{(5)}R)$ theory, as it is usually done in common 4D $f(R)$ theories.\\

Once we have the field equations of the $f(^{(5)}\! R)$ theo\-ry, we are now in position to study its stability. 

\subsection{Dolgov-Kawasaki instability in 5D}

 In order to derive a Dolgov-Kawasaki instability criterion on this $f(^{(5)}\!R)$ gravity theory let us to use the parametrization 
\begin{equation}\label{p4}
f(^{(5)}\!R)=\,^{(5)}\!R+\gamma\zeta(^{(5)}\!R),
\end{equation}
where $\gamma$ is a small parameter with $[length]^{-2}$ units. This election of $f(^{(5)}\!R)$ means that we are considering deviations of the theory from 5D general relativity. Inserting (\ref{p4}) in (\ref{p3}) and evaluating the 5D D'Alambertian we obtain
\begin{eqnarray}
&& ^{(5)}\Box\,^{(5)}\!R+\frac{\zeta^{(3)}}{\zeta^{(2)}}\nabla^{a}\,\!^{(5)}\!R\nabla_{a}\,\!^{(5)}\!R+\frac{\left(\gamma\zeta^{(1)}-\frac{5}{2}\right)\,^{(5)}\!R}{4\gamma\zeta^{(2)}}\nonumber \\
\label{p5}
&& =\frac{\kappa_5\,^{(5)}T}{4\gamma\zeta^{(2)}}+\frac{5\zeta}{8\zeta^{(2)}},
\end{eqnarray}
where $\zeta^{(i)}$ is denoting the higher derivative of order $i$ with respect to $^{(5)}\!R$. We are also assuming that $\zeta^{(2)}\neq 0$ to avoid the 5D general relativity case. Following a similar procedure to that employed in \cite{Faraoni1}, we will consider the weak field limit conditions
\begin{equation}\label{p6}
g_{ab}=\eta_{ab}+H_{ab},\quad ^{(5)}\!R=-\kappa_{5}\,^{(5)}T+\,\!^{(5)}\!R_1,
\end{equation}
where $\eta_{ab}$ is the 5D Minkowsky metric, $H_{ab}$ is a 5D me\-tric fluctuation tensor respect to the Minkowsky background, $^{(5)}\!R_1$ is a first order perturbation to $^{(5)}\!R$ and $|\,^{(5)}\!R/\kappa_5\,^{(5)}T|\ll 1$ with $^{(5)}T\neq 0$. Thus, linearizing the equation (\ref{p5}) it leads to
\begin{eqnarray}
&& ^{(5)}\!\ddot{R}_1-\nabla^{2}\,\!^{(5)}\!R_1-\,^{(5)}\!\overset{\star\star}{R}_1-\frac{2\kappa_5\zeta^{(3)}}{\zeta^{(2)}}\,\!^{(5)}\dot{T}\,\!^{(5)}\!\dot{R}_1\nonumber \\
&&
+\frac{2\kappa_5\zeta^{(3)}}{\zeta^{(2)}}\nabla\,\!^{(5)}T\cdot\nabla\,\!^{(5)}R_1+\frac{2\kappa_5\zeta^{(3)}}{\zeta^{(2)}}\,\!^{(5)}\overset{\star}{T}\,\!^{(5)}\overset{\star}{R}_1\nonumber \\
&& +\frac{1}{4\zeta^{(2)}}\left[\frac{5}{2\gamma}-\zeta^{(1)}\right]\,\!^{(5)}\!R_1 = \frac{1}{4\zeta^{(2)}}\left[\kappa_5\left(\frac{7}{2\gamma}-\zeta^{(1)}\right)\,\!^{(5)}T\right.\nonumber \\
\label{p7}
&&\left. +\frac{5}{2}\zeta\right]+\kappa_5\,\!^{(5)}\ddot{T}-\kappa_5\nabla^{2}\,\!^{(5)}T-\kappa_{5}\,\!^{(5)}\overset{\star\star}{T},
\end{eqnarray}
where the dot denotes time derivative, the operator $\nabla^2$ is the 3D Laplacian operator and the star $(\star)$ indicates derivative with respect to the fifth extended extra dimension $l$. 
It can be easily seen from (\ref{p7}) that given the smallness of $\gamma$ the dominant contribution in the effective mass term (the coefficient of $^{(5)}\!R_1$) is $8\gamma\zeta^{(2)}$ and hence, as occurs in the usual 4D case, the stability condition in 5D continues being $f_{,RR}>0$.\\

In braneworld scenarios sources of matter in 5D are usually regarded, even in some $f(^{(5)}\! R)$ braneworld mo\-dels \cite{Bfr}. However, in theories like the induced matter approach of P.S. Wesson no matter in 5D is considered, so they assume a 5D vacuum. Thus, for these kind of cases we have that in the absense of matter or in the presence of traceless matter $^{(5)}\!T=0$, the equation (\ref{p5}) yields the linearized expression
\begin{eqnarray}
&& ^{(5)}\!\ddot{R}_1-\,^{(5)}\!\overset{\star\star}{R}_1+\frac{\zeta^{(3)}}{\zeta^{(2)}}\,^{(5)}\!\dot{R}_1^2-\frac{\zeta^{(3)}}{\zeta^{(2)}}\,^{(5)}\!\overset{\star}{R}_1^2-\nabla^2\,^{(5)}\!R_1\nonumber\\
&&-\frac{\zeta^{(3)}}{\zeta^{(2)}}(\nabla\,^{(5)}R_1)^2+\frac{1}{4\zeta^{(2)}}\left[\frac{5}{2\gamma}-\zeta^{(1)}\right]\,^{(5)}\!R_1=\frac{5\zeta}{8\zeta^{(2)}}.\nonumber \\
\label{vip1}
\end{eqnarray}
Again in the effective mass term the dominant contribution comes from $8\gamma\zeta^{(2)}$, and thus the criterion to avoid a negative effective mass remains: $f_{,RR}>0$, under the presence of traceless 5D matter sources.

\subsection{Field equations of both some braneworlds and induced matter theory as maximally symmetric solutions}

Maximally symmetric solutions are very common in  $f(R)$ theories of gravity. Due to the Jebsen-Birkoff theorem in 4D the Schwarzschild solution is no more unique in this kind of theories \cite{Faraoni2}. In fact, when we go up from 4D to 5D in a theory of the type of general relativity, the Birkoff theorem is no more valid \cite{wbook}. Thus, we expect to find more spherically symmetric solutions in a $f(\,\!^{(5)}\!R)$ theory of gravity than on its analog in 4D.\\

With this idea in mind let us study ma\-xi\-ma\-lly symme\-tric solutions in the theory prescribed by the action (\ref{p1}). As it is well-known a maximally symmetric solution is characterized by a constant Ricci scalar, in this case by $^{(5)}\!R=\,\!^{(5)}\!R_{0}$. Hence the trace expression (\ref{p3}) yields
\begin{equation}\label{p8}
 f_{,R}(^{(5)}\!R_0)\,^{(5)}\!R_0 - \frac{5}{2} f(^{(5)}\!R_0) =\kappa_5\,^{(5)}T.
\end{equation}
The field equations (\ref{p2}) for constant scalar curvature spaces reduces to
\begin{equation}\label{p9}
f_{,R}(^{(5)}\!R_0)\,^{(5)}\!R_{ab}- \frac{1}{2} f(^{(5)}\!R_0) g_{ab}=\kappa_5 ^{(5)}T_{ab}.
\end{equation}
A combination of (\ref{p8}) and (\ref{p9}) leads to
\begin{equation}\label{p10}
^{(5)}\!R_{ab}=\frac{\kappa_5\,\!^{(5)}\!R_0\,\!^{(5)}\!T_{ab}+\frac{1}{2}f(\,\!^{(5)}\!R_0)\,\!^{(5)}\!R_0g_{ab}}{\kappa_5\,\!^{(5)}\!T+\frac{5}{2}f(\,\!^{(5)}R_0)}.
\end{equation}
For traceless 5D matter fields, the equation (\ref{p10}) leads to
\begin{equation}\label{p12}
^{(5)}\!R_{ab}=\kappa_{eff5}\,\!^{(5)}\!T_{ab}+\frac{1}{5}\,\!^{(5)}\!R_0g_{ab},
\end{equation}
where $\kappa_{eff5}=[2\kappa_5\,\!^{(5)}\!R_0]/[5f(\,\!^{(5)}\!R_0)]$. These are the field equations of  braneworld scenarios with traceless 5D energy-momentum tensor and a 5D cosmological constant term. \\
In the absence of matter sources $^{(5)}T_{ab}=0$, the expression (\ref{p10}) becomes
\begin{equation}\label{p12a}
^{(5)}\! R_{ab}=\frac{1}{5}\,^{(5)}\! R_0 g_{ab},
\end{equation}
which for $^{(5)}\!R_0=0$ correspond to the 5D field equations of the induced matter theory of gravity \cite{Wesson}. When  $^{(5)}\!R_0>0$ it describes a De-Sitter spacetime, but when $^{(5)}\!R_0<0$ this space corresponds to an Anti-De-Sitter spacetime, which is the one employed for example in Randall-Sundrum models \cite{Re7,Re8}.\\

 In summary, we can say in a fashion that both the field equations of the induced matter theory and the ones of some braneworld models can be obtained from a $f(^{(5)}\! R)$ theory of gravity as particular maximally symmetric solutions.

\section{The induced 4D field equations}

We are now in position to derive the 4D field equations induced from the 5D dynamics. In order to do so, we choose a 5D coordinate chart in which the 5D line element can be written as
\begin{equation}\label{p13}
dS_{5}^2=g_{\alpha\beta}(x^{\sigma},l)dx^{\alpha}dx^{\beta}+\epsilon \Phi^{2}(x^{\sigma},l)dl^2,
\end{equation}
where $\epsilon=\pm 1$ accounts for the signature of the extra non-compact coordinate $l$ and $\Phi(x^{\sigma},l)$ is a well-behaved metric function.\\

Now, let us assume that the 5D spacetime can be foliated by a family of hypersurfaces, generically defined by $\Sigma_{l}:l=l(x^{\mu})$. Hence, we can consider for example a cons\-tant foliation $\Sigma_0:l=l_0$ or a dynamical one $\Sigma_t:l=l(t)$. Thus, the 4D line element induced on every hypersurface $\Sigma_l$ is given by
\begin{equation}\label{p14}
dS_{4}^2=h_{\mu\nu}(x^{\sigma})dx^{\mu}dx^{\nu},
\end{equation}
where $h_{\mu\nu}(x^{\sigma})=\left.g_{\mu\nu}(x^{\sigma},l)\right|_{\Sigma_l}$ is the 4D induced metric tensor.\\
 Some useful quantities and operators in order to implement the dimensional reduction from 5D to 4D, can be expressed in their (4+1) form as 
\begin{eqnarray}\label{p15}
\nabla_{\mu}\nabla_{\nu}f_{,R}&=& {\cal D}_\mu{\cal D}_\nu f_{,R} +\frac{\epsilon}{2\Phi^2}\overset{\star}{g}_{\mu\nu}\overset{\star}{f_{,R}},\\
\label{p16}
\nabla_{l}\nabla_{l}f_{,R}&=& \overset{\star\star}{f_{,R}}+\epsilon\Phi\left({\cal D}_\alpha\Phi\right)\left({\cal D}^\alpha f_{,R}\,\right)-\frac{\overset{\star}{\Phi}}{\Phi} \overset{\star}{f_{,R}},\\
^{(5)}\!\Box f_{,R} & =&\Box f_{,R}-\frac{\left({\cal D}_\alpha \Phi\right)\left({\cal D}^\alpha f_{,R}\right)}{\Phi}\nonumber \\
\label{p17}
&+&\frac{\epsilon}{\Phi^2}\left[\overset{\star\star}{f_{,R}}+\left(\frac{1}{2}g^{\mu\nu}\overset{\star}{g}_{\mu\nu}-\frac{\overset{\star}{\Phi}}{\Phi} \right)\overset{\star}{f_{,R}}\right],\\
^{(5)}\!R_{\mu\nu}  &=& R_{\mu\nu}-\frac{{\cal D}_\mu{\cal D}_\nu \Phi}{\Phi}+\frac{\epsilon}{2\Phi^2}\left(\frac{\overset{\star}{\Phi}}{\Phi} \overset{\star}{g}_{\mu\nu}-\overset{\star\star}{g}_{\mu\nu}\right.\nonumber \\
\label{p18}
&&\left.+{g}^{\lambda\sigma}\,\overset{\star}{g}_{\mu\lambda}\,\overset{\star}{g}_{\sigma\nu}-\frac{1}{2}{g}^{\alpha\beta}\overset{\star}{g}_{\alpha\beta}\overset{\star}{g}_{\mu\nu}\right),\\
^{(5)}\!R_{ll}& = &  -\epsilon\Phi\,\Box\Phi-\frac{1}{4}\overset{\star}{g}^{\alpha\beta}\overset{\star}{g}_{\alpha\beta}-\frac{1}{2}{g}^{\alpha\beta}\overset{\star\star}{g}_{\alpha\beta}+\frac{1}{2}\frac{\overset{\star}{\Phi}}{\Phi}g^{\alpha\beta}\overset{\star}{g}_{\alpha\beta},\nonumber \\
\label{p19}
\end{eqnarray}
where ${\cal D}_{\mu}$ is denoting the 4D covariant derivative and $\Box=h^{\mu\nu}{\cal D}_{\mu}{\cal D}_{\nu}$ is the 4D D'Alambertian operator. Thus, with the help of (\ref{p15}) to (\ref{p19}) the $\mu\nu$ and $ll$ components of the field equations (\ref{p2}) can be combined to obtain on 4D hypersurfaces
%\begin{small}
\begin{eqnarray}
&&\left. \left[f_{,R}(^{(5)}\!R)\,R_{\mu\nu} - \frac{1}{2} f(^{(5)}\!R) g_{\mu\nu} - ({\cal D}_\mu{\cal D}_\nu-g_{\mu\nu}\,\Box)f_{,R}(^{(5)}\!R)\right]\right|_{\Sigma_l}\nonumber \\
\label{p20}
&&=\left.\kappa_5\,^{(5)}T_{\mu\nu}\right|_{\Sigma_l}+\kappa_4f_{,R}(^{(5)}\!R)\tau_{\mu\nu},
\end{eqnarray}
%\end{small}
where the matter tensor $\tau_{\mu\nu}$ is defined by \cite{wbook,rept,Wesson}
\begin{eqnarray}
&& \kappa_4 \tau_{\mu\nu} = \kappa_4 T_{\mu\nu}^{(IM)}+ \frac{\epsilon}{2\Phi^2}\,\frac{\overset{\star}{g}_{\mu\nu}\overset{\star}{f}_{,R}}{f_{,R}}-
g_{\mu\nu}\frac{\left({\cal D}_\alpha \Phi\right)\left({\cal D}^\alpha f_{,R}\right)}{\Phi f_{,R}}\nonumber\\
&&-\frac{\epsilon\, g_{\mu\nu}}{\Phi^2}\left[\frac{\overset{\star\star}{f}_{,R}}{f_{,R}}+\left(\frac{1}{2}g^{\alpha\beta}\overset{\star}{g}_{\alpha\beta}-\frac{\overset{\star}{\Phi}}{\Phi} \right)\frac{\overset{\star}{f}_{,R}}{f_{,R}}\right] 
  -\frac{1}{4}{g}_{\mu\nu}\left[\overset{\star}{g}^{\lambda\sigma}\overset{\star}{g}_{\lambda\sigma}\right.\nonumber\\
\label{p21}
&&\left.+\left(g^{\lambda\sigma}\overset{\star}{g}_{\lambda\sigma}\right)^2\right]
\end{eqnarray}
being $T_{\alpha\beta}^{(IM)}$ the energy-momentum tensor for geo\-me\-tri\-ca\-lly induced matter, that was first introduced in the Wesson's induced matter theory \cite{rept}, which is given by
\begin{eqnarray}
&&\kappa_4T_{\mu\nu}^{(IM)}=\frac{h_{\mu\nu}\Box\Phi}{\Phi}-\frac{\epsilon}{2\Phi^{2}}\left\lbrace \frac{\overset{\star}{\Phi}}{\Phi}\overset{\star}{h}_{\mu\nu}-\overset{\star\star}{h}_{\mu\nu}+h^{\lambda\alpha}\overset{\star}{h}_{\mu\lambda}\overset{\star}{h}_{\nu\alpha}\right.\nonumber\\
\label{p22}
&&\left.-\frac{1}{2}h^{\alpha\beta}\overset{\star}{h}_{\alpha\beta}\overset{\star}{h}_{\mu\nu}
+\frac{1}{4}h_{\mu\nu}\left[\overset{\star}{h}^{\alpha\beta}\overset{\star}{h}_{\alpha\beta}+(h^{\alpha\beta}\overset{\star}{h}_{\alpha\beta})^{2}\right]\right\rbrace.
\end{eqnarray}
In order to evaluate the $f(^{(5)}\! R)$ terms in (\ref{p20}) on the 4D hypersurface $\Sigma_l$, we express the 5D Ricci scalar curvature as a function of its analog 4D in the form \cite{wbook,rept,Wesson}
\begin{eqnarray}
&&^{(5)}\!R=\,^{(4)}\!R-\frac{\Box\Phi}{\Phi}+\frac{\epsilon}{2\Phi^2}\left(\frac{\overset{\star}{\Phi}}{\Phi}g^{\mu\nu}\overset{\star}{g}_{\mu\nu}-g^{\mu\nu}\overset{\star\star}{g}_{\mu\nu}\right.\nonumber\\
\label{p23}
&&\left. +g^{\mu\nu}g^{\lambda\alpha}\overset{\star}{g}_{\mu\lambda}\overset{\star}{g}_{\alpha\nu}-\frac{1}{2}g^{\mu\nu}g^{\alpha\beta}\overset{\star}{g}_{\alpha\beta}\overset{\star}{g}_{\mu\nu}\right),
\end{eqnarray}
which can also be written in terms of the extrinsic curvature tensor $K_{\alpha\beta}$ as: $^{(5)}\!R=\,^{(4)}\!R-(K^{\mu\nu}K_{\mu\nu}-K^2)$, with $K=h^{\alpha\beta}k_{\alpha\beta}$. Thus, using (\ref{p23}) the field equations (\ref{p20}) on the 4D spacetime $\Sigma_l$ read
\begin{eqnarray}
&&F^{\prime}(\,^{(4)}\!R,\varphi)R_{\mu\nu}-\frac{1}{2}F(\,^{(4)}\!R,\varphi)h_{\mu\nu}-\left[{\cal D}_{\mu}{\cal D}_{\nu}\right.\nonumber\\
&&\left. -h_{\mu\nu}\Box\right]F^{\prime}(\,^{(4)}\!R,\varphi)=\kappa_4 S_{\mu\nu}+\kappa_{4}F^{\prime}(\,^{(4)}\!R,\varphi)\tau_{\mu\nu},\nonumber\\
\label{p24}
\end{eqnarray}
where the prime is denoting derivative with respect to $^{(4)}\!R$, the function $F(\,\!^{(4)}\!R,\varphi)=f[\,\!^{(5)}\!R=\,\!^{(4)}\!R+E]|_{\Sigma_l}$ is the induced function of the 4D Ricci scalar, $\kappa_4S_{\mu\nu}=[\kappa_5 (T_{\mu\nu}-g_{\mu\nu}(\,^{(5)}\!T_{ll}-(\epsilon\Phi^{2}/3)\,^{(5)}\!T))]|_{\Sigma_l}$  and $\varphi(x^{\sigma})=E(x^{\sigma},l)|_{\Sigma_l}$, being the extrinsic scalar curvature parameter $E$ defined by
\begin{eqnarray}
&&E(x^a)=-\frac{\Box\Phi}{\Phi}+\frac{\epsilon}{2\Phi^2}\left(\frac{\overset{\star}{\Phi}}{\Phi}g^{\mu\nu}\overset{\star}{g}_{\mu\nu}-g^{\mu\nu}\overset{\star\star}{g}_{\mu\nu}\right.\nonumber\\
\label{p25}
&&\left.+g^{\mu\nu}g^{\lambda\alpha}\overset{\star}{g}_{\mu\lambda}\overset{\star}{g}_{\alpha\nu}-\frac{1}{2}g^{\mu\nu}g^{\alpha\beta}\overset{\star}{g}_{\alpha\beta}\overset{\star}{g}_{\mu\nu}\right).
\end{eqnarray}
The field equations (\ref{p24}) describe a new type of $F(\,\!^{(4)}\!R,\varphi)$ theory where a matter coupling of the form $F^{\prime}\tau_{\mu\nu}$ is present. When the parameter $E(x^{a})$ depends only of the extra coordinate, the effective scalar field $\varphi$ becomes a constant, and in this case the $F(\,\!^{(4)}\!R,\varphi)$ transforms into a $F(^{(4)}\!R)$ theory with the same matter coupling. As we will see in a forthcoming example, when the parameter $E$ has only time dependence, the $F(^{(4)}\!R,\varphi)$ resulting theory can be interpreted as an effective modified gravity theory with dynamical coefficients. If a 5D vacuum is considered, then the sources of matter are exclusively induced by the 5D geometry of the theory. Notice that the modified gravity theory described by the equations (\ref{p24}) is in fact an effective theory, in the sense that it is derived from the 5D one described by the action (\ref{p1}).\\

For example, for the line element
\begin{equation}\label{ejemplo1}
ds_5^{2}=\left(\frac{l}{l_0}\right)^{2\alpha}\left[dt^2-a^{2}(t)\delta_{ij}dx^{i}dx^{j}\right]-\Phi(t)^2dl^2,
\end{equation}
the extrinsic scalar curvature parameter $E(x^{a})$ reads
\begin{equation}\label{ejemplo2}
E(t)=2\left[\frac{\ddot{\Phi}}{\Phi}+3H\frac{\dot{\Phi}}{\Phi}-\frac{2\alpha (5\alpha-2)}{l^2\Phi^2}\right].
\end{equation}
Thus, on our 4D spacetime $\Sigma_0:l=l_0$ the effective scalar field is given by
\begin{equation}\label{ejemplo3}
\varphi(t)=2\left[\frac{\ddot{\Phi}}{\Phi}+3H\frac{\dot{\Phi}}{\Phi}-\frac{2\alpha (5\alpha-2)}{l_0^2\Phi^2}\right].
\end{equation}
Hence, if we consider for example a $f(^{(5)}\!R)=^{(5)}\!R+\frac{b(t)}{^{(5)}\!R}$, in 4D the induced effective $F(^{(4)}\!R,\varphi)$ has the form
\begin{equation}\label{ejemplo4}
F(^{(4)}\!R,\varphi)=\,^{(5)}\!R+\varphi(t)+\frac{b(t)}{^{(4)}\!R+\varphi(t)}.
\end{equation}
This $F(^{(4)}\!R,\varphi)$ can be interpreted as a $F(^{(4)}\!R)$ modified gravity theory with dynamical coefficients, similar to the one proposed in \cite{Re9}. As it was shown in \cite{Re9}, this kind of models are very useful to describe the dark energy dominance epoch in the evolution of the universe. A remarkable difference between the models in \cite{Re9} and the example given by (\ref{ejemplo4}) relies in the form of the field equations. Moreover, in here the scalar field $\varphi$ is not a dynamical coefficient introduced apriori, instead it is determined by the 5D geometry. \\

Now, let us to give another application. If we consider for example a 5D line element of a warped product spacetime
\begin{equation}\label{ejemplo5}
  ds_5^{2}=e^{2A(l)}\left[dt^2-a^2(t)\delta_{ij}dx^{i}dx^{j}\right]-dl^2,
\end{equation}
in this case the extrinsic scalar curvature parameter $E(x^{a})$ results
\begin{equation}\label{ejemplo6}
E(l)=-4\left(2\overset{\star\star}{A}+5\overset{\star}{A}^2\right).
\end{equation}
Evaluating (\ref{ejemplo6}) on the 4D hypersurface $\Sigma_0:l=l_0$, the effective scalar field has the expression
\begin{equation}\label{ejemplo7}
\varphi(l_0)=-\left.4\left[2\overset{\star\star}{A}+5\overset{\star}{A}^2\right]\right|_{l=l_0},
\end{equation}
which clearly is a constant. Therefore, the effective $F(^{(4)}\!R,\varphi)$ effective theory becomes a matter coupled $F(^{(4)}\!R)$ modified gravity theory in this case.\\

Now, returning to the non-diagonal components of the field equations (\ref{p2}), the components $\mu l$ of the field equations (\ref{p2}) can be written as
\begin{small}
\begin{equation}
  \label{conserv1}
  (\Phi f_{,R}){\cal D}_\alpha {\cal P}_\mu^{\,\alpha}=\kappa_5\!^{(5)}T_{\mu l}+\frac{1}{2}g^{\alpha\sigma}\overset{\star}{g}_{\sigma\mu}f_{,RR}R_{,\alpha}-\frac{\Phi_{,\mu}}{\Phi}\overset{\star}{f}_{,R}+\overset{\star}{f}_{,RR}R_{,\mu}
\end{equation}
\end{small}
 where
\begin{equation}
  \label{conserv2}
{\cal P}_{\alpha\beta}=\frac{1}{2\Phi}\left(\overset{\star}{g_{\alpha\beta}}-g_{\alpha\beta}g^{\mu\nu}\overset{\star}{g_{\mu\nu}}\right).
\end{equation}
A similar equation to (\ref{conserv1}) is obtained in the induced ma\-tter theory of gravity \cite{Wesson}. In that theory the analogous expression is  
\begin{equation}
  \label{conserv3}
{\cal D}_{\alpha}{\cal P}_\mu^{\,\alpha}=0.
\end{equation}
The conservation like equation (\ref{conserv3}) can be recovered from (\ref{conserv1}) when we consider  $f(^{(5)}\!R)=\, ^{(5)}\!R$ in vacuum i.e. without any sources of matter in 5D.

\subsection{Dolgov-Kawasaki instability criterion for the effective $F(\,\!^{(4)}\!R,\varphi)$ theory}

In order to study the Dolgov-Kawasaki instability in the matter sector of the effective $F(\,\!^{(4)}\!R,\varphi)$ theory induced from a $f(^{(5)}\! R)$ gravity theory, we will proceed as follows.\\

The trace of the field equations (\ref{p24}) leads to
\begin{equation}\label{inesta1}
3\Box F^{\prime}+F^{\prime}\,^{(4)}\!R-2F=\kappa_4 S+\kappa_4 F^{\prime}\tau,
\end{equation}
being $S=h^{\mu\nu}S_{\mu\nu}$ and $\tau=h^{\mu\nu}\tau_{\mu\nu}$. Now, deviations from Einstein's general relativity of our $F(\,\!^{(4)}\!R,\varphi)$ are described by the expression
\begin{equation}\label{inesta2}
F(\,\!^{(4)}\!R,\varphi)=^{(4)}\!R+\sigma Z(^{(4)}R,\varphi),
\end{equation}
where $\sigma$ is a small parameter with $[length]^{-2}$ units. Employing (\ref{inesta2}) the trace equation (\ref{inesta1}) yields
\begin{eqnarray}
&&\Box\,^{(4)}\!R+\frac{Z^{\prime\prime\prime}}{Z^{\prime}}{\cal D}^{\mu}\,^{(4)}\!R{\cal D}_{\mu}\,^{(4)}\!R+\frac{1}{Z^{\prime\prime}}\frac{\partial^2 Z^{\prime}}{\partial\varphi^2}{\cal D}^{\mu}\varphi{\cal D}_{\mu}\varphi\nonumber \\
&&+\frac{3\sigma}{Z^{\prime\prime}}\frac{\partial Z^{\prime}}{\partial\varphi}\Box\varphi +\frac{(\sigma Z^{\prime}-2)\,^{(4)}\!R}{3\sigma Z^{\prime\prime}}=\kappa_4\frac{(S+\sigma Z^{\prime}\tau)}{3\sigma Z^{\prime\prime}}+\frac{2Z}{3Z^{\prime\prime}}.\nonumber\\
\label{inesta3}
\end{eqnarray}
In the weak field regime we can use the approximation
\begin{equation}\label{inesta4}
h_{\alpha\beta}=\eta_{\alpha\beta}+\gamma_{\alpha\beta},\quad ^{(4)}\!R=R_b+R_1,
\end{equation}
where $\eta_{\alpha\beta}$ is the 4D Minkowsky metric, $\gamma_{\alpha\beta}$ is a fluc\-tua\-tion of the metric with respect to the Minkowsky background, $|R_1/R_b|\ll 1$ and $R_b=-\kappa_4(S+F^{\prime}\tau)$. Using (\ref{inesta4}), the equation (\ref{inesta3}), to first order in $R_1$, reads
\begin{eqnarray}
&& \ddot{R}_1-\nabla^2R_1-\frac{2\kappa_4 Z^{\prime\prime\prime}}{Z^{\prime\prime}}\dot{S}\dot{R}_1-\frac{2\kappa_4 Z^{\prime\prime\prime}}{Z^{\prime\prime}}\left(\dot{Z}^{\prime}\tau +Z^{\prime}\dot{\tau}\right)\dot{R}_1\nonumber\\
&& +\frac{2\kappa_4 Z^{\prime\prime\prime}}{Z^{\prime\prime}}\nabla S\cdot\nabla R_1 +\frac{2\kappa_4 Z^{\prime\prime\prime}}{Z^{\prime\prime}}\left(\nabla Z^{\prime}\tau +Z^{\prime}\nabla\tau\right)\cdot\nabla R_1\nonumber\\
&&+\frac{1}{3Z^{\prime\prime}}\left(\frac{1}{\sigma}-Z^{\prime}\right)R_1+\frac{1}{Z^{\prime\prime}}\frac{\partial^2 Z^{\prime}}{\partial\varphi^2}{\cal D}^{\mu}\varphi{\cal D}_{\mu}\varphi + \frac{3\sigma}{Z^{\prime\prime}}\frac{\partial Z^{\prime}}{\partial\varphi}\Box \varphi\nonumber\\
&& =\kappa_4\ddot{S}+\kappa_4\sigma\left(\ddot{Z}^{\prime}\tau +2\dot{Z}^{\prime}\dot{\tau}+Z^{\prime}\ddot{\tau}\right)-\kappa_4\left[\nabla^2 S\right. \nonumber\\
\label{inesta5}
&&\left. +\nabla^2(\sigma Z^{\prime}\tau)\right]-\frac{\kappa_4\left(S+\sigma Z^{\prime}\tau\right)Z^{\prime}}{3Z^{\prime\prime}}-\frac{Z}{3Z^{\prime\prime}}.
\end{eqnarray}
Clearly the effective mass term is dominated by the factor $3\sigma Z^{\prime\prime}$ and then the theory is  stable only if $F^{\prime\prime}>0$.

\section{Final Comments}

In this letter we have discussed some implications of considering a 5D gravity governed by a $f(^{(5)}\!R)$ theory on spacetimes with a non-compact spacelike fifth extra coordinate. In this theoretical setting our 4D universe is described by a generic hypersurface $\Sigma_l$, embedded into the 5D spacetime. Applying a dimensional reduction mechanism we obtain on $\Sigma_l$ a set of induced 4D field equations that describe a $F(^{(4)}\!R,\varphi)$ modified gravity theory which exhibits a matter coupling term of the form: $\kappa_{4}F^{\prime}(\,^{(4)}\!R,\varphi)\tau_{\mu\nu}$. This coupling is different from the one in BBHL theory. If we consider a 5D vacuum ($^{(5)}T_{ab}=0$), matter sources in our 4D universe are induced geometrically by the 5D geometry in a similar manner as it is done in the Wesson's induced matter theory. \\

In general in a BBHL theory, in order to recover the Einstein-Hilbert action it is necessary to specify the two functions of the scalar curvature as: $f_{1}(R)=R$ and $f_2(R)=1$. In our case it is sufficient to fix $f(^{(5)}\!R)=\,^{(5)}\!R$ on a warped product metric background for example and automatically $F(^{(4)}\!R,\varphi)=\,^{(4)}\!R+\varphi_0$ and $F^{\prime}(^{(4)}\!R,\varphi)=1$, resulting the field equations (\ref{p24}) in the general relativity equations. Thus the matter coupling term is governed by the same $F(^{(4)}\!R,\varphi)$, instead to fix two different functions as in the case of BBHL theory. The Dolgov-Kawasaki instability criterion of both the $f(^{(5)}\!R)$ theory and the effective $F(^{(4)}\!R,\varphi)$ continus being the same than in usual $f(R)$ theories: $f_{,RR}(^{(5)}\!R)>0$ and $F^{\prime\prime}(^{(4)}\!R,\varphi)>0$. \\

When the effective scalar field $\varphi$ becomes only time dependent, the resulting $F(^{(4)}\!R,\varphi)$ theory can be interpreted as a modified gravity theory with dynamical coe\-ffi\-cients. The main difference with respect to this kind of models in the literature relies on it dynamical equations (\ref{p24}). According to \cite{Re9}, this kind of models may be viable to explain the present accelerated expansion of the universe. The study of this cosmological solutions of the effective $F(^{(4)}\!R,\varphi)$ theory will be matter of future work.

\section{Acknowledgements}

\noindent J.E.M.A acknowledges CONACYT M\'exico, Centro Universitario de Ciencias Exactas e Ingenierias and Centro Universitario de los Valles, of Universidad de Guadalajara for financial support.

\end{document}